\documentclass[cameraready]{Interspeech}

\title{Quantifying the Uncertainty of Blindly Estimated Room Embeddings \\ Using a Dispersion-Calibrated Score}

\author[affiliation={1},orcid=0009-0001-2239-4632]{Yang}{Xiang}
\author[affiliation={2}]{Philipp}{G{\"o}tz}
\author[affiliation={2},orcid=0000-0002-2613-8046]{Emanu\"{e}l A.\,P.}{Habets}
\author[affiliation={3}]{Andreas}{Walther}
\author[affiliation={1}]{Wenwu}{Wang}
\author[affiliation={1},orcid=0000-0001-7933-5935]{Philip~J.\,B.}{Jackson}

\address{
    $^1$ Centre for Vision Speech and Signal Processing, University of Surrey, United Kingdom \\
        $^2$ International Audio Laboratories Erlangen\textsuperscript{$\ast$}, Erlangen, Germany\thanks{\textsuperscript{$\ast$}A joint institution of Fraunhofer IIS and Friedrich-Alexander-Universit{\"a}t Erlangen-N{\"u}rnberg (FAU), Germany.}
    \\
     $^3$ Fraunhofer Institute for Integrated Circuits IIS, Erlangen, Germany 
}

\email{yang.xiang@surrey.ac.uk, p.jackson@surrey.ac.uk}


\usepackage{comment}
\usepackage[sort]{cite}
\usepackage{adjustbox}

\begin{document}

\maketitle

\begin{abstract}

Room embeddings derived from reverberant speech are often unreliable: speech content and recording degradation can alter the representation even when speaker, room, and source–receiver geometry remain unchanged, degrading downstream task performance. We propose a framework that learns room embeddings robust to speech-content variation
and a representation-level uncertainty score from reverberant speech without downstream-task supervision. The embedding is anchored to a structured room impulse response (RIR) latent space and trained using a multi-view data structure with Kullback--Leibler (KL)-based alignment; a multi-positive contrastive term further refines robustness. A lightweight uncertainty head is calibrated using the dispersion of corruption-induced embeddings and optimized with a rank-based objective. Across waveform- and spectrogram-level corruptions, the score is consistent with representation dispersion and enables effective selective prediction while requiring only a single utterance at inference.

\end{abstract}

\section{Introduction}
Reverberant speech carries cues about the acoustic environment, enabling applications such as blind estimation of physical parameters (e.g., reverberation time $T_{60}$ and clarity index $C_{50}$)~\cite{ratnam2003blind,eaton2016estimation,looney2020joint}, acoustic-environment retrieval/verification~\cite{khokhlov2024classification} and robust speech processing~\cite{kinoshita2016summary}.
Existing approaches broadly fall into two directions: (i) task-specific blind inference, including room impulse response (RIR) reconstruction~\cite{steinmetz2021filtered,liao2023blind,lee2023yet} and acoustic-parameter estimation, and (ii) task-agnostic representation learning that yields reusable room embeddings~\cite{khokhlov2019r,gotz2024blind,gotz2026multi}, which can be transferred across downstream objectives.

However, learning \emph{reliable} room embeddings remains challenging because the observed signal is confounded by non-room factors (e.g., content, speaker, noise). In practice, recordings are rarely clean: noise and partial signal loss (e.g., dropouts, bandwidth limits) can substantially alter embeddings even when the room and source--receiver geometry are unchanged, degrading representation quality and downstream robustness. This motivates a \emph{general-purpose} uncertainty score that indicates whether an embedding should be trusted. While uncertainty estimation has been widely studied in classification and regression \cite{kendall2017uncertainties,gal2016dropout}, recent blind room-acoustic inference work has begun to quantify uncertainty by calibrating error bounds for RIR/parameter estimates derived from learned environment representations \cite{gotz2026multi}. Yet, predicting task-agnostic uncertainty scores for \emph{representation reliability} with controlled speech corruptions is underexplored.

Following \cite{gotz2026multi}, we learn a task-agnostic room representation from reverberant speech together with an uncertainty score reflecting representation reliability under corruption. Our approach has three stages. First, we learn a structured room-acoustic latent space using a variational autoencoder (VAE) trained on log-mel RIR spectrograms. Second, we train a speech encoder with multi-view training and contrastive learning to improve robustness to speech-content variation while remaining aligned with the RIR latent space. Third, we train a lightweight uncertainty head that estimates representation uncertainty from embedding dispersion under controlled corruptions. Compared with \cite{gotz2026multi}, Stage-1 uses the same RIR-VAE latent space, while Stage-2 adds multi-view KL-based alignment with a multi-positive contrastive term, and Stage-3 converts corruption-induced dispersion into a task-agnostic, single-utterance uncertainty score.

Our main contributions are:
(1)~We identify the multi-view data structure (multiple speech realizations per RIR) as a key driver of robustness to speech-content variation under latent alignment.
(2)~We propose a general-purpose room embedding framework that combines KL-based alignment to a frozen RIR-VAE latent space with a multi-positive contrastive term for a modest verification gain.
(3)~We introduce a dispersion-calibrated uncertainty score, trained with a rank-based objective to map corruption-induced embedding dispersion to a single-utterance uncertainty estimate.

We evaluate representation quality on RIR verification, RIR log-mel reconstruction, and acoustic-parameter prediction, and assess reliability using uncertainty--dispersion correlation and selective prediction on verification and reconstruction. Overall, the score detects unreliable embeddings under the considered corruptions.


\section{Problem Formulation}
We model the observed reverberant speech signal $y[t]$ as 
an anechoic speech source $x[t]$ convolved with an RIR $h[\ell]$ of length $L$, corrupted by additive noise $w[t]$:
\begin{equation}
y[t] = \sum_{\ell=0}^{L-1} h[\ell]\,x[t-\ell] + w[t],
\label{eq:conv_model}
\end{equation}
where $t$ is the discrete-time sample index and $\ell$ indexes the RIR taps.
 In this work, both reverberant speech and RIRs are represented by
log-mel spectrograms $\mathbf{Y}\in\mathbb{R}^{F\times T}$ and $\mathbf{H}\in\mathbb{R}^{F\times T}$, with $F$  frequency bands and $T$ time frames.

Given $\mathbf{Y}$ only (blind setting), our goal is to learn a task-agnostic room representation $\mathbf{z}_Y=f_\theta(\mathbf{Y})$ that captures room acoustics while being robust to speech-content variation, and to estimate a scalar uncertainty $U=g_\psi(\mathbf{z}_Y)$ that reflects representation reliability under varying acoustic conditions.

\section{Method}
\begin{figure}[!t]
  \centering
  \includegraphics[width=0.47\textwidth]{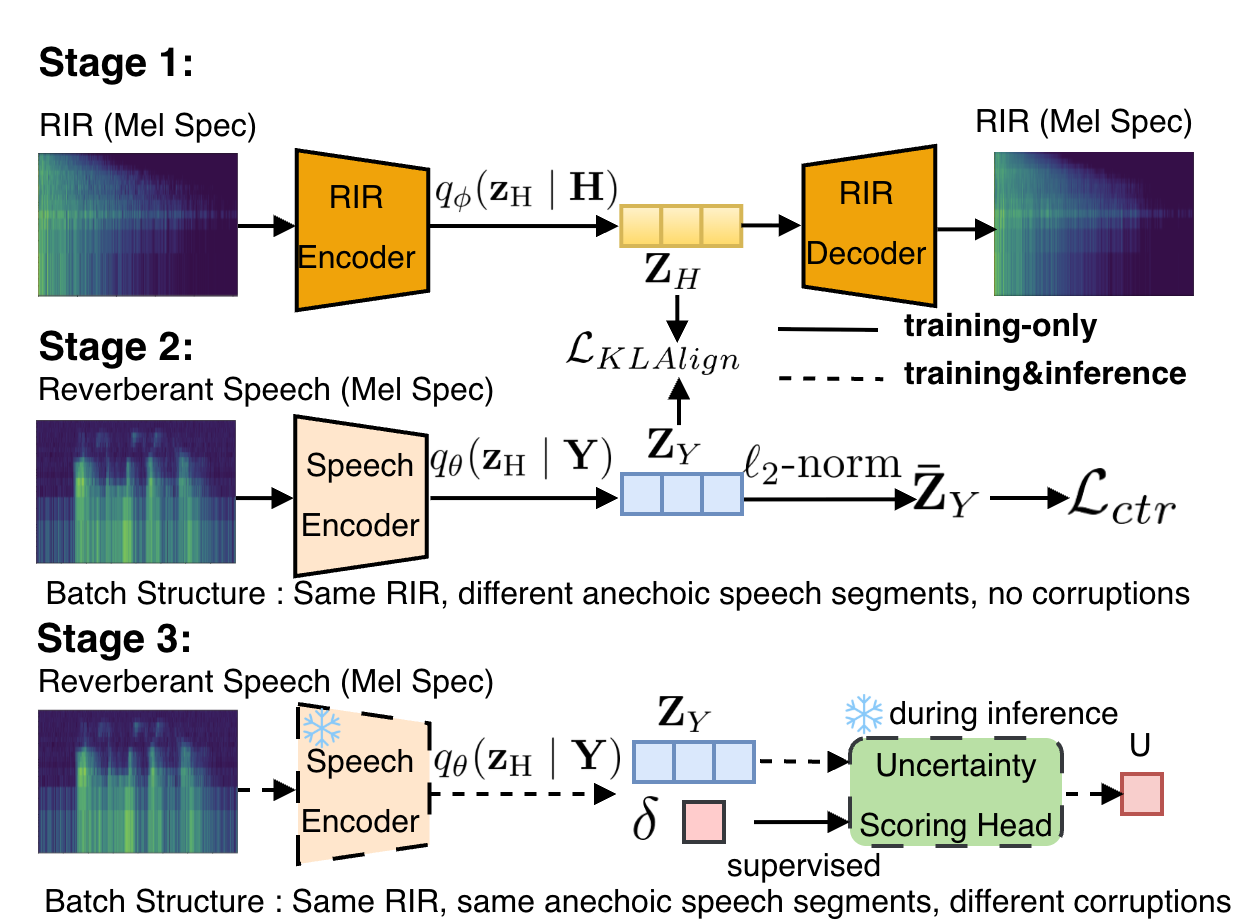}
  \caption{Overview of the three-stage training pipeline. Stage-1 trains an RIR-VAE to obtain posterior statistics $(\boldsymbol{\mu}_{z_H}, \boldsymbol{\sigma}^2_{z_H})$. Stage-2 learns a speech embedding $\mathbf{z}_Y$ using multi-positive contrastive learning and KL alignment to the frozen RIR posterior. Stage-3 freezes the speech encoder and trains a scalar uncertainty head supervised by embedding dispersion $\delta$.}
  \label{fig:framework}
\end{figure}

Figure~\ref{fig:framework} summarizes our three-stage training pipeline. Stages~1--2 follow the latent-approximation anchoring strategy in~\cite{gotz2026multi}, while we introduce (i) multi-positive contrastive learning to improve robustness to speech-content variation and (ii) a dispersion-supervised uncertainty head for predicting task-agnostic representation reliability.



\subsection{Stage-1: RIR-VAE Pretraining}
We train a VAE~\cite{kingma2014aevb} on RIR log-mel spectrograms $\mathbf{H}\in\mathbb{R}^{F\times T}$ to obtain a structured latent space. The encoder outputs a diagonal-Gaussian posterior
$q_\phi(\mathbf{z}_H \mid \mathbf{H})=\mathcal{N}\left(\boldsymbol{\mu}_{z_H}, \mathrm{diag}(\boldsymbol{\sigma}^2_{z_H})\right)$,
where $\boldsymbol{\mu}_{z_H}$ and $\boldsymbol{\sigma}^2_{z_H}$ have the same shape as $\mathbf{z}_H$.
Specifically, we parameterize $\mathbf{z}_H$ as a latent tensor of size $C\times F'\times T' = 64\times 4\times 16$, where $(F',T')$ are the frequency and time resolutions after the encoder’s strided downsampling of the input spectrogram.
The VAE is trained with an $\ell_2$ reconstruction loss ($\widehat{\mathbf{H}}$ denotes the reconstructed RIR log-mel spectrogram) and a KL regularizer:
\begin{equation}
\mathcal{L}_{\text{vae}}=\lVert \widehat{\mathbf{H}}-\mathbf{H} \rVert_2^2+\lambda_{1}\,\mathrm{KL}\left(q_\phi(\mathbf{z}_H\mid \mathbf{H})\,\Vert\,p(\mathbf{z}_H)\right).
\end{equation}
\subsection{Stage-2: Contrastive Embedding with RIR Alignment}


{\textbf{Speech Encoder Architecture.}} The speech encoder \cite{gotz2026multi} adopts a hybrid CNN--Transformer design, using a strided 2D CNN front-end to encode log-mel spectrograms and a 3-layer Transformer with attention pooling~\cite{vaswani2017attention} to produce an utterance-level embedding $\mathbf{z}_Y\in\mathbb{R}^{D}$. We set $D=4096$ to match the flattened RIR-VAE latent dimensionality for KL-based alignment.
%
%
%
%
We use cosine similarity for contrastive learning and dispersion; thus we compute an $\ell_2$-normalized embedding $\bar{\mathbf{z}}_Y = \mathbf{z}_Y/\lVert \mathbf{z}_Y\rVert_2$. The encoder outputs raw $\mathbf{z}_Y\in\mathbb{R}^D$, which is used for KL alignment and the uncertainty head.



{\textbf{RIR Latent Alignment.}}
This alignment serves as a space regularizer using RIR-only self-supervision (Stage-1) and does not use any downstream task labels or heads.
To anchor speech embeddings to the RIR latent space, we align the raw speech embedding $\mathbf{z}_Y$ to the frozen RIR-VAE posterior
$q_\phi(\mathbf{z}_H \mid \mathbf{H})=\mathcal{N}(\boldsymbol{\mu}_{z_H},\mathrm{diag}(\boldsymbol{\sigma}^2_{z_H}))$ via a KL term.
Following \cite{gotz2024blind,gotz2026multi}, we treat $\mathbf{z}_Y$ as the mean of a Gaussian with fixed identity covariance $\mathbf{I}$,
$q_\theta(\mathbf{z}_H \mid \mathbf{Y}) =\mathcal{N}(\mathbf{z}_Y, \mathbf{I})$.
We fix the covariance to $\mathbf{I}$ to avoid introducing an additional uncertainty model in Stage-2; the goal is geometric anchoring rather than probabilistic inference.
We minimize
\begin{equation}
\mathcal{L}_{\text{align}} 
=
\mathrm{KL}\left(
q_\theta(\mathbf{z}_H \mid \mathbf{Y})\ \Vert\ q_\phi(\mathbf{z}_H \mid \mathbf{H})
\right).
\end{equation}
For diagonal Gaussians, the KL has a closed form; we use the mean-over-dim implementation as in~\cite{gotz2026multi}.

{\textbf{Contrastive Objective.}}
To improve robustness to speech-content variation,  we propose to employ a multi-positive contrastive learning approach~\cite{gotz2023contrastive,khosla2020supervised} where positive pairs share the same RIR:
\begin{equation}
\begin{aligned}
\mathcal{L}_{\mathrm{ctr}}
&=
-\frac{1}{N}
\sum_{n:\,|P(n)|>0}
\frac{1}{|P(n)|}
\sum_{m\in P(n)}
\log
\frac{\exp(s_{nm})}
{\sum_{k \ne n}\exp(s_{nk})},\\
P(n)
&=
\{\, m \ne n \mid \mathrm{id}_m = \mathrm{id}_n \,\},
\qquad
s_{nm}=\frac{\bar{\mathbf{z}}_{Y,n}^\top \bar{\mathbf{z}}_{Y,m}}{\tau}.
\end{aligned}
\end{equation}
%
%
Here, $N$ is the batch size after flattening, $\tau$ is a temperature parameter,
and $P(n)$ denotes the set of positives in the batch sharing the same RIR
label ($\mathrm{id}$) as anchor $n$. The loss averages uniformly over all positives for
each anchor; anchors without positives are excluded. This objective requires a multi-view batch construction. 

The total loss used to train Stage-2 is
\begin{equation}
\mathcal{L}_{\text{encoder}}
=
\lambda_2\,\mathcal{L}_{\text{ctr}}
+
(1-\lambda_2)\,\mathcal{L}_{\text{align}},
\qquad \lambda_2 \in [0,1].
\end{equation}
The total loss encourages embeddings to remain discriminative across RIRs while being geometrically consistent with the RIR latent space. This training is supervised by RIR IDs (contrastive + latent alignment) and does not depend on any downstream objective, yielding a task-agnostic representation $\mathbf{z}_Y$.

\subsection{Stage-3: Dispersion-Calibrated Uncertainty}
In Stage-3, we freeze the encoder and train a lightweight uncertainty head $U=g_\psi(\mathbf{z}_Y)$ to predict the dispersion, without any downstream supervision; $U$ is thus task-agnostic. The head is a two-layer multilayer perceptron with a single hidden layer (size 256) with a Softplus output, ensuring $U\ge0$.

{\textbf{Embedding Dispersion.}}
A ``view'' $v$ denotes one corrupted version of the same utterance (noise, frequency mask, or time mask), paired with its uncorrupted clean anchor.
For each corrupted sample, dispersion is defined relative to its clean anchor
using cosine distance in the $\ell_2$-normalized embedding space:
\begin{equation}
\label{dispersion equation}
\delta_v = 1 - \cos\left(\bar{\mathbf{z}}_{Y,v}, \bar{\mathbf{z}}_{Y,\text{clean}}\right).
\end{equation}

{\textbf{Training Objective (Rank Loss).}}
In Stage-3, we train the uncertainty head to produce scores that are
\emph{order-consistent} with the corruption-induced representation shift,
rather than matching its absolute scale. 
We form pairs $(v_1,v_2)$ of corrupted samples (optionally restricted to
the same corruption type). If $\delta_{v_1} > \delta_{v_2}$, we enforce $U_{v_1} > U_{v_2}$ by a
margin-based ranking loss:
\begin{equation}
\mathcal{L}_{\text{rank}}
= \mathbb{E}_{(v_1,v_2)} \left[ \max\bigl(0,\; \gamma - (U_{v_1} - U_{v_2})\bigr) \right],
\end{equation}
where $\gamma$ is a margin. This objective learns a monotonic mapping from
dispersion to uncertainty, 
avoiding sensitivity to $\delta$'s absolute
magnitude. Early stopping is selected by maximizing the Spearman
correlation between $U$ and $\delta$ on corrupted samples.

\section{Experimental Setup}
\subsection{Datasets}
To generate reverberant speech, anechoic utterances were taken from the EARS dataset \cite{richter2024ears}. 
A curated collection of measured RIRs was constructed from several 
publicly available datasets, including the ACE Challenge dataset \cite{eaton2016estimation}, AIR-IKS \cite{jeub2009binaural}, ASH-IR \cite{ash_ir_dataset}, Arni \cite{arni_dataset_2022}, BUT ReverbDB \cite{8717722}, EchoThief \cite{echothief}, GTU-RIR \cite{gtu_rir_dataset}, MIT IR Survey \cite{traer2016statistics}, Multi-Purpose RIR Dataset\cite{Friede2024multi-purpose}, Multi-Room Transition dataset \cite{gotz2022neural}, MeshRIR \cite{koyama2021meshrir}, Motus \cite{gotz2021dataset}, OpenAIR \cite{shelley2010openair}, SoundCam\cite{wang2023soundcam}, SurrRoom\cite{cieciuraSurrRoomDatasetSpatial2023}, TAU-SRIR \cite{politis2022_tausrir}, and TH Köln dataset \cite{lubeck2021high}. 

In total, 3000 RIRs ($T_{60}\in[0.046,\,1.898]\,\mathrm{s}$) were collected and randomly partitioned into three mutually exclusive subsets for training, validation, and testing. We split by RIR identity (each measured source--receiver position pair defined one RIR); a room could contribute multiple RIRs at different positions, but no RIR appeared in more than one split. An identical split strategy was applied to all anechoic speech signals. Based on these splits, approximately 89~hours of reverberant speech were generated for training, 11~hours for validation, and 11~hours for testing, using 
4-s segments derived from the respective subsets.
For Stage-2 training, we used \emph{multi-view} (MV) batches (256 samples): 16 RIRs $\times$ 16 utterances. \emph{Single-view} (SV) batches, as used in~\cite{gotz2026multi},  contain one utterance per RIR (256 RIRs $\times$ 1; 256 distinct utterances).
All RIRs were converted to log-mel spectrograms with 16 mel bands, computed at 16 kHz using a short-time Fourier transform (STFT) with a window length of 64 samples and a hop size of 16 samples (75\% overlap). 
The reverberant speech representations followed the same configuration, except for a hop size of 32 samples (50\% overlap).

\subsection{Experimental Settings}
\textbf{Speech Corruptions.} To probe embedding robustness under varying acoustic conditions, we applied controlled corruptions at waveform and spectrogram levels to the reverberant speech signals. 
We added waveform-domain pink noise with signal-to-noise ratio (SNR) sampled over speech-active regions: 15 to 25\,dB (\textit{mild}), 5 to 15\,dB (\textit{medium}), $-$5 to 5\,dB (\textit{severe}). On log-mel spectrograms, we applied SpecAugment-style frequency and time masking \cite{park2019specaugment}. Frequency masking zeroed $k_f=\mathrm{round}(\rho_f F)$ contiguous mel bands, with $\rho_f$ sampled from 5--15\% (\textit{mild}), 15--25\% (\textit{medium}), or 25--35\% (\textit{severe}). Time masking zeroed $k_t=\mathrm{round}(\rho_t T)$ contiguous frames (sequence length preserved), with $\rho_t$ sampled from 0--10\% (\textit{mild}), 10--20\% (\textit{medium}), or 20--30\% (\textit{severe}). The frequency- and time-masking ratios were inspired by the augmentation policies of SpecAugment \cite{park2019specaugment}, extended here up to 35\% and 30\%, respectively.
The corruption severity was drawn with probabilities 0.7, 0.25, and 0.05 for \textit{mild}, \textit{medium}, and \textit{severe}, respectively, yielding progressively stronger but mostly mild perturbations for dispersion and uncertainty analysis. 

\noindent \textbf{Training hyperparameters.} Stage-1 uses $\lambda_{1}=0.05$; Stage-2 uses $\tau=0.1$ and $\lambda_2=1/13$; Stage-3 uses $\gamma=0.1$.

\subsection{Acoustic Parameters}
For evaluation, we computed acoustic parameters from the \emph{ground-truth} RIRs and used the values as reference labels.
All parameters were computed in $B$\,=\,$7$ octave bands, implemented via a 2nd-order Butterworth filterbank, with center frequencies
$f_c^{(b)}$\,$\in$\,$\{125, 250, 500, 1\mathrm{k}, 2\mathrm{k}, 4\mathrm{k}, 8\mathrm{k}\}\,\mathrm{Hz}$, indexed by $b$\,$\in$\,$\{1,\dots,B\}$\@.
$T_{60}$ was obtained from a DNN-based decay-slope estimator \cite{gotz2022neural}, and $C_{50}$ from the band-limited RIR energy ratio as  
\begin{equation}
C_{50}^{(b)} = 10 \log_{10}
\left(
\frac{\sum_{t=0}^{t_{50}-1} h_{(b)}^{2}[t]}
{\sum_{t=t_{50}}^{L-1} h_{(b)}^{2}[t]}
\right),
\end{equation}
where $h_{(b)}[t]$ denotes the band-limited RIR in the $b$-th octave band (length $L$) at time index $t$, and $t_{50}$ is the sample index corresponding to 50\,ms after the direct-path arrival.

\subsection{Performance Metrics}
We evaluated (i) the representation quality of Stage-2 embeddings, and (ii) the reliability of the learned uncertainty score $U$. For representation quality, we report RIR verification~\cite{khokhlov2024classification} (rather than room verification, to assess discrimination across positions within the same room) performance via pair-based retrieval (Average Precision, AP) and reconstruction error of the frozen RIR decoder in the log-mel spectrogram domain (MAE in dB)~\cite{gotz2024blind,gotz2026multi}. For blind acoustic parameter estimation, we report errors for $T_{60}$ and $C_{50}$ (macro-averaged across the 7 standard octave bands). 

We also evaluated uncertainty, i.e. how well $U$ reflects corruption-induced representation shift via Spearman correlation between $U$ and embedding dispersion $\delta$. Last, we evaluated \emph{selective prediction}: with samples sorted by $U$, we show downstream performance with respect to retained coverage.

\section{Experimental Results}
We report the results for stage-2 representation quality and uncertainty-dispersion consistency as detailed below. 

\subsection{Stage-2 Representation Quality}
\label{sec:exp_stage2}

Table~\ref{tab:stage2_main} summarizes Stage-2 representation quality on (a) RIR verification, (b) RIR reconstruction via the frozen RIR-VAE decoder, and (c) blind estimation of $T_{60}$ and $C_{50}$. 
For verification, we measure \emph{Average Precision} (AP, area under the precision--recall curve) computed from pairwise cosine similarities, where positives correspond to pairs sharing the same RIR identity and negatives to different RIRs. This metric is designed to reflect the consistency of room embeddings from the same RIR across different speech content. 
MAE$_{\text{rec}}$ denotes the mean absolute error between the decoded and ground-truth log-mel magnitudes, in~dB. MAPE$_{T_{60}}$ denotes the mean absolute percentage error for $T_{60}$.

We compare against FiNS~\cite{steinmetz2021filtered} (time-domain blind RIR estimation from reverberant speech; for reconstruction, its estimated RIR is converted to log-mel magnitudes) and the baseline of G{\"o}tz \emph{et al.}~\cite{gotz2026multi}, referred to as MRL (a multi-stage representation learning pipeline with conformal quantile uncertainty).
In our setting, MRL-SV is trained with the SV data construction of~\cite{gotz2026multi}, while MRL-MV keeps the same MRL architecture and objective but replaces SV with our MV data construction; we treat MRL-MV as a data-construction ablation.

\begin{table}[t]
  \caption{Stage-2 representation quality and task performance.}
  \label{tab:stage2_main}
  \centering
    \footnotesize
  \setlength{\tabcolsep}{3pt} 
  \begin{adjustbox}{width=0.45 \textwidth}
  \begin{tabular}{@{}lcccc@{}}
    \toprule
    \textbf{Method} &
    \textbf{AP} $\uparrow$ &
    \textbf{MAE}$_{\text{rec}}$\,(dB) $\downarrow$ &
    \textbf{MAPE}$_{T_{60}}$\,(\%) $\downarrow$ &
    \textbf{MAE}$_{C_{50}}$\,(dB) $\downarrow$ \\
    \midrule
    FiNS \cite{steinmetz2021filtered} & 0.82 & 9.59 & 29.08 & 2.90 \\
    MRL-SV (KL)  \cite{gotz2026multi} & 0.95 & 4.76 & 16.67 & 1.90 \\
    MRL-MV (KL) & 0.98 & \textbf{4.04} & 12.87 & \textbf{1.49} \\
    Proposed (Ctr+KL)& \textbf{0.99} & 4.06 & \textbf{12.86} & 1.50 \\
    \bottomrule
  \end{tabular}
  \end{adjustbox}
\end{table}
Compared to MRL-SV, our MRL-MV ablation improves verification (AP: 0.98 vs.\ 0.95),  supporting the view that MV speech realizations per RIR improve robustness to speech-content variation under latent alignment. Building on this setting, our proposed model further adds an explicit multi-positive contrastive term, yielding a modest additional gain for verification (AP: 0.99 vs.\ 0.98). 
Across reconstruction and blind parameter estimation, results remain similar between MRL-MV and Proposed, suggesting that the contrastive term primarily provides a refinement for verification/retrieval without materially changing performance on the other tasks in Table~\ref{tab:stage2_main}.


\subsection{Uncertainty--Dispersion Consistency}
\label{sec:exp_ud}

We evaluate whether an uncertainty score is \emph{order-consistent} with corruption-induced representation dispersion, quantified by the clean-anchored dispersion $\delta$ (Eq.~\ref{dispersion equation}), using Spearman's rank correlation $\rho(U,\delta)$ (monotonic and scale-free). Importantly, clean anchors are only needed to compute $\delta$ (and for training), while inference predicts $U(\mathbf{Y})$ from a single corrupted utterance. We report type-wise correlations for additive noise, frequency masking (F.Mask), and time masking (T.Mask). We include two reference scores: \textbf{Severity}, which uses the corruption control parameter (noise: SNR; masking: ratio), and \textbf{MRL-MV}, which uses the conformally calibrated inter-quantile range of subband RIR reconstruction errors as uncertainty (retrained under the same corruption protocol). Note that a \emph{global} correlation across corruption types is not reported for Severity, since the control parameters are not comparable across corruption types.

Table~\ref{tab:spearman_udelta} shows strong uncertainty--dispersion consistency for our method (global $\rho=0.90$; robust across corruption types), whereas Severity correlates weakly with dispersion and MRL-MV correlates less strongly than ours, indicating that coarse corruption controls and reconstruction-error uncertainty reflect representation-level instability less reliably under our corruptions.

\begin{table}[t]
\centering
\caption{Spearman correlation $\rho(U,\delta)$ between uncertainty score $U$ and corruption-induced representation dispersion $\delta$.}
\label{tab:spearman_udelta}
\footnotesize
\begin{tabular}{lcccc}
\hline
\textbf{Method} & \textbf{Global} $\uparrow$ & \textbf{Noise} $\uparrow$ & \textbf{F.Mask} $\uparrow$ & \textbf{T.Mask} $\uparrow$ \\
\hline
Severity & - & 0.28 & 0.16 & 0.17 \\
MRL-MV & 0.85 & 0.59 & 0.66 & 0.68 \\
Proposed ($U$)      & \textbf{0.90}       & \textbf{0.83}       & \textbf{0.79}       & \textbf{0.86} \\
\hline
\end{tabular}
\end{table}

\subsection{Downstream Reliability via Selective Prediction}
\label{sec:exp_selective}
We assess whether uncertainty $U$ can act as a downstream uncertainty score via \emph{selective prediction}: we rank all corrupted samples by a criterion, retain the top-$c$ fraction with the \emph{lowest} predicted risk, and report performance as a function of coverage $c\in(0,1]$. $U$ is computed from the corrupted utterance only.

We compare two ranking criteria. \textbf{$U$-sorted} ranks samples in ascending order of $U$ (low $\rightarrow$ high), i.e., we keep samples with the lowest uncertainty first. As an oracle baseline that uses corruption metadata, \textbf{severity-sorted} ranks samples by the corruption control parameter (noise: descending SNR; masking: ascending mask ratio), i.e., cleaner inputs first. Under both criteria, decreasing coverage removes increasingly corrupted samples; hence, the left side of the curve (small coverage) corresponds to retaining only the cleanest inputs.

\begin{figure}[t]
\centering
\includegraphics[width=0.95\linewidth]{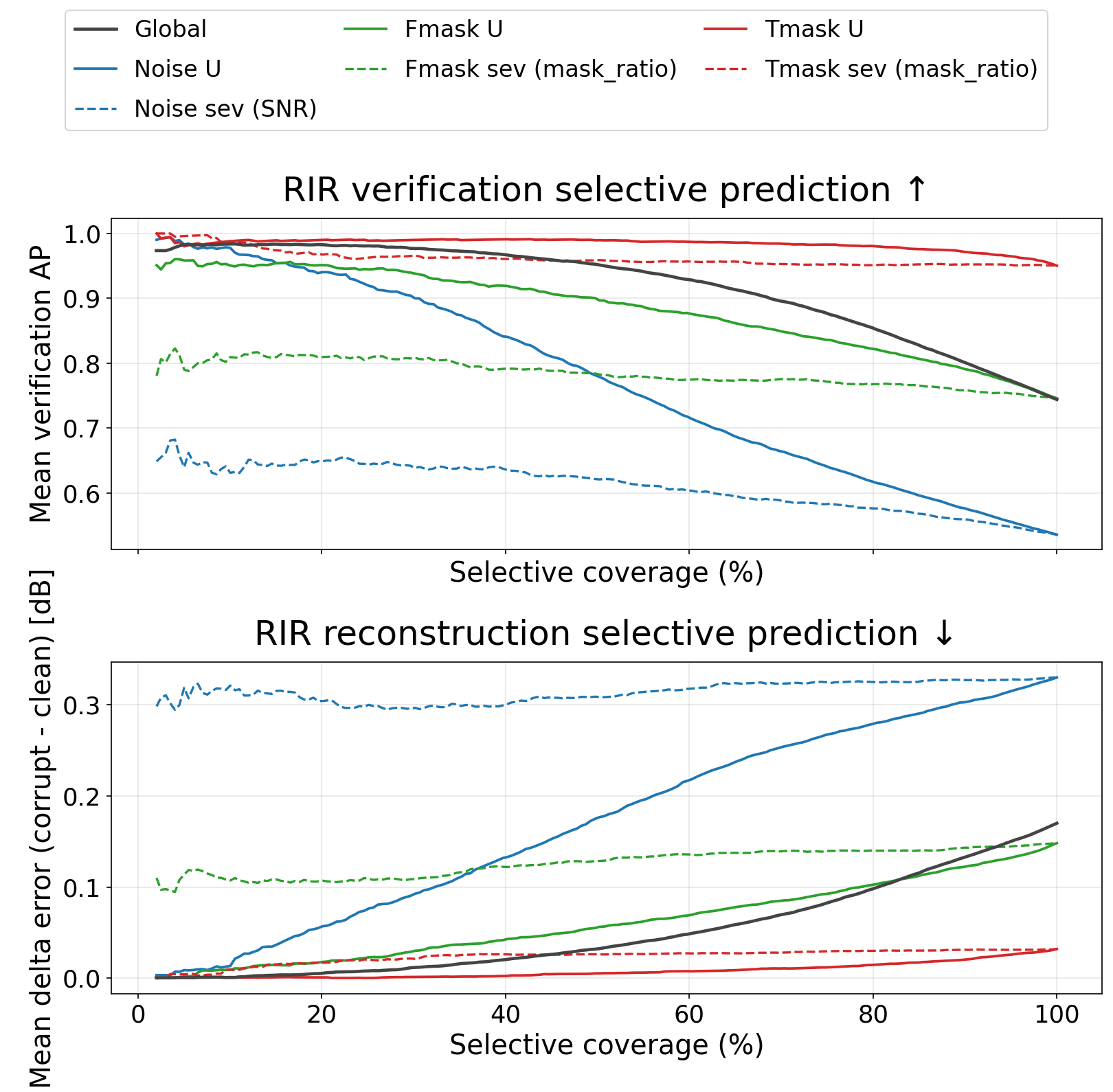}
\caption{Selective prediction. Samples are sorted by uncertainty
$U$ (solid) or by corruption severity controls (dashed).}
\label{fig:selective_verif}
\end{figure}

Figure~\ref{fig:selective_verif} shows selective curves for RIR verification AP and reconstruction $\Delta$ (corrupt--clean), stratified by corruption type. Across both evaluations, $U$-sorting yields a steadier improvement as coverage decreases than severity-sorting, indicating that $U$ provides a finer-grained reliability ranking than the coarse corruption controls. This suggests that downstream degradation is not fully explained by the control parameter alone, whereas $U$ better captures the induced representation shift and enables more effective filtering of unreliable samples.

\section{Conclusion and Limitations}
Building on~\cite{gotz2026multi}, we improve robustness to speech-content variation in room embeddings
learned from reverberant speech and learn a task-agnostic uncertainty score. 
Our 3-stage pipeline (i) learns a structured RIR latent space using an RIR-VAE, (ii) trains a speech encoder using a multi-view data structure with KL-based alignment to the RIR latent space, adding a multi-positive contrastive term for a modest additional verification gain, and (iii) trains a lightweight uncertainty head via a rank loss to predict corruption-induced embedding dispersion.
Experiments under controlled waveform- and spectrogram-level corruptions show that $U$ correlates strongly with representation shift and enables effective selective prediction with single-utterance inference. 

While the proposed score supports single-utterance inference and is validated via downstream selective prediction, limitations include that $U$ is a dispersion-calibrated uncertainty score (not a posterior uncertainty) and Stage-3 training relies on clean-anchored dispersion targets (paired clean/corrupted views); our split is by RIR identity (not strictly room-disjoint); and the corruption set (pink noise, SpecAugment masks) does not cover in-the-wild degradations (e.g., interfering speakers, device mismatch, clipping).  

\section{Generative AI Use Disclosure}
Generative AI tools were used only for language editing and polishing (e.g., improving clarity, grammar, and style). All scientific content, including the technical contributions, experimental design, results, and conclusions, was produced and verified by the authors. All authors take full responsibility for the content of this paper and consent to its submission.

\bibliographystyle{IEEEtran}
\bibliography{mybib}

\end{document}